\documentclass{IEEEtran4PSCC}

\usepackage{amssymb}
\usepackage{algorithm}
\usepackage{etoolbox}
\usepackage{algpseudocode}

\usepackage{tablefootnote}

\ifCLASSINFOpdf
   \usepackage[pdftex]{graphicx}
\else
   \usepackage[dvips]{graphicx}
\fi
\usepackage[cmex10]{amsmath}
\usepackage[dvipsnames]{xcolor}
\usepackage{array}
\usepackage{booktabs}
\usepackage{bm}
\usepackage{algorithm,algpseudocode}
\usepackage{fancyhdr}

\newtheorem{remark}{\bf Remark}

\newcommand{\mppse}[1]{$\mathbf{{P}^{MP}_{cktPSE}}$}
\newcommand{\mcpse}[1]{$\mathbf{{P}^{MC}_{cktPSE}}$}
\newcommand{\nppse}[1]{$\mathbf{{P}_{cktPSE}}$}
\newcommand{\sbtpse}[1]{$\mathbf{P^{SBT}_{cktPSE}}$}

\algdef{SE}[SUBALG]{Indent}{EndIndent}{}{\algorithmicend\ }%
\algtext*{Indent}
\algtext*{EndIndent}

\makeatletter
\let\old@ps@headings\ps@headings
\let\old@ps@IEEEtitlepagestyle\ps@IEEEtitlepagestyle
\def\psccfooter#1{%
    \def\ps@headings{%
        \old@ps@headings%
        \def\@oddfoot{\strut\hfill#1\hfill\strut}%
        \def\@evenfoot{\strut\hfill#1\hfill\strut}%
    }%
    \def\ps@IEEEtitlepagestyle{%
        \old@ps@IEEEtitlepagestyle%
        \def\@oddfoot{\strut\hfill#1\hfill\strut}%
        \def\@evenfoot{\strut\hfill#1\hfill\strut}%
    }%
    \ps@headings%
}
\makeatother

\psccfooter{%
        \parbox{\textwidth}{\hrulefill \\ \small{To be presented at \\
        XXIII Power Systems Computation Conference} \hfill \begin{minipage}{0.2\textwidth}\centering \vspace*{4pt} \includegraphics[scale=0.06]{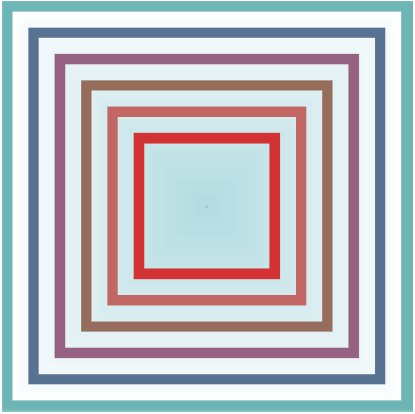}\\\small{PSCC 2024} \end{minipage} \hfill \small{Paris, France --- June 4 -- 7, 2024}}%
}

\begin{document}
%
\title{Circuit-theoretic Joint Parameter-State Estimation --- Balancing Optimality and AC Feasibility}

\author{
\IEEEauthorblockN{Peng Sang, and Amritanshu Pandey}
\IEEEauthorblockA{ Dept. of Electrical and Biomedical Engineering\\
The University of Vermont\\
Burlington, VT USA\\
\{Peng.Sang, Amritanshu.Pandey\}@uvm.edu}

}


\maketitle

\begin{abstract}
AC State Estimation (ACSE) is widely recognized as a practical approach for determining the grid states in steady-state conditions. It serves as a fundamental analysis to ensure grid security and is a reference for market dispatch. As grid complexity increases with rapid electrification and decarbonization, there is a growing need for more accurate \textit{knowledge} of the grid operating state. However, existing ACSE algorithms have technical gaps. Critically, current ACSE algorithms are susceptible to \textit{erroneous} system parameters, which are assumed to be fixed in traditional approaches. In this paper, we build a novel circuit-theoretic joint parameter-state estimation algorithm to address this limitation. The innovative algorithm builds an analogous equivalent circuit of the grid with states and certain parameters \textit{unknown}. It solves a circuit-constrained optimization to estimate the \textit{most likely} grid states and parameters given a set of measurements. Further, it quantifies the goodness of the estimated output by formulating \textit{tight} convex envelopes around the original non-convex problem to quantify the quality of estimates. We compare the various proposed approaches on systems with up to 2869 nodes while demonstrating a tradeoff between solution optimality and model fidelity.
\end{abstract}


\thanksto{\noindent This material is based upon work supported by the U.S. Department of Energy's Office of Energy Efficiency and Renewable Energy (EERE) under the Solar Energy Technologies Office Award Number DE-EE0010147.The views expressed herein do not necessarily represent the views of the U.S.Department of Energy or the United States Government.}

\section*{Nomenclature}

\addcontentsline{toc}{section}{Nomenclature}
\begin{IEEEdescription}[\IEEEusemathlabelsep\IEEEsetlabelwidth{$V_1,V_2,V_3$}]

\small 
\item[\textbf{Symbol}] \textbf{Interpretation}
\item[$V_r^k,V_i^k$] Real/imag voltage at bus $k$
\item[$I_r^k,I_i^K$] Real/imag current injection at bus $k$
\item[$V^{kl}_r,V^{kl}_i$] Real/imag voltage across bus $k$ and $l$
\item[$|V^k|,\theta^k$] Voltage magnitude/angle at bus $k$
\item[$P^k,Q^k$] Active/reactive power injection at bus $k$
\item [\textbf{z}] Set of measurements
\item[$z^k_{|V|}, z^k_P, z^k_Q$] Remote terminal units (RTU) injection measurements of voltage mag., real/reactive power at bus $k$
\item[$z^{kl}_{P,fl}, z^{kl}_{Q,fl}$] RTU flow measurements for a edge between $k$ and $l$;real/reactive power at the metered end
\item [$z^{met}_{|V|}$] RTU flow measurements for a edge between $k$ and $l$;voltage mag. at the metered end
\item [\textbf{v}] Set of voltage state variables 
\item[$v^{k}$] Voltage pair at bus $k$; $x^{k} = [V^{k}_r, V^{k}_i]$, $\in$ \textbf{v}
\item[$n^k$] Noise term (in per unit system)
\item[$n^k_r,n^k_i$] Real and imag. noise term at bus k 
\item[$n^k_{|V|}$] Voltage mag. noise term at bus k
\item[$I(v) = 0$] Network balance Kirchoff's Current Law (KCL) equations
\item[$I_r^{kl},I_i^{kl}$] Real/imag current injection from bus $k$ to $l$
\item[$W$] Error covariance matrix
\item[$w_I^k, w_V^k$] Weights for current and voltage measurements at bus $k$, respectively.
\item[$G^{kl},B^{kl}$] Admittance terms between bus $k$ and $l$
\item[$B_{sh}^{k}$] Shunt susceptance at bus $k$
\item [$\boldsymbol{\mathcal{P}}$] Set of unknown parameters
\item [$\boldsymbol{\mathcal{P}^U}$, $\boldsymbol{\mathcal{P}^L}$] Set of original upper and lower bounds of unknown parameters 
\item [$\boldsymbol{\check{\mathcal{P}}^U}$, 
$\boldsymbol{\mathcal{\check{P}}^L}$] Set of tight upper and lower bounds of unknown parameters 
\item [\textbf{x}] Set of variables, \textbf{x} = $\boldsymbol{\mathcal{P}}$ $\cup$ \textbf{v}

\item [$\mathcal{I}$] Set of injection measurements

\item [$\mathcal{E}$] Set of flow measurements

\item [$\mathcal{R}$] Set of RTU measurements, $\mathcal{R} = \mathcal{I} \cup \mathcal{E}$
\end{IEEEdescription}

\section{Introduction}


\subsection{Motivation of the research}


\noindent The outputs from the network topology processor (NTP) and AC state-estimator (ACSE) construct real-time models of power grids in control rooms. 
These models are critical to evaluate the grid reliability through what-if-like contingency analysis and also serve as base networks for downstream real-time economic dispatch.
Amongst other things, the quality of these grid models is very sensitive to the value of the input model parameters.
If the model parameter values are not sufficiently accurate, the ACSE solution and the quality of the downstream models deteriorate.
These parameter values include but are not limited to impedances of the transmission lines and transformers, shunt values, and transformer tap positions.


There exist many scenarios today when the network parameter values are erroneous or unknown and undetectable by general bad-data flow. 
These scenarios include equipment aging, temperature variation, unreported conductor reconfiguration, loss of a single conductor in a parallel configuration, unreported transformer tap adjustments, and manual database entry errors. 
In most of these scenarios, inaccurate backlogged parameters will be fed into the ACSE, with potentially damaging impact.


Grid control rooms have processes to detect bad data, including erroneous measurements and bad parameters.
The most common is a post-processing approach where operators apply statistical tests on ACSE output (i.e., the residual) to identify suspect measurements and parameter candidates.
While the approach is effective in the presence of only gross bad-measurement data, there is evidence that wrong parameters do not always show up in residual output \cite{Abur2004}.
Localization is challenging too with such statistical approaches. 
The largest residual value does not always correspond to localized error information \cite{Merrill1971}.
Also, distinguishing measurement from parameter errors remains challenging, one being a random variable and the other an unknown quantity \cite{Zanni2020}.




\subsection{State-of-the-art and limitations}

\noindent Alternatively, erroneous parameters can be assumed unknown and obtained via a joint parameter-state estimation problem, which can estimate both state and unknown parameters concurrently.
The approach is also called “state (vector) augmentation” and has been widely studied for both transmission and distribution networks \cite{Abur2004}, \cite{Jovicic2020}, \cite{Vanin2023}. 
The approach overcomes some challenges with the post-processing approach based on ACSE residuals, but other challenges persist.
The original joint approach \cite{Abur2004} formulates the joint problem as a weighted least square problem \cite{Schweppe1970} and applies the Gauss-Newton solution approach.
The problem is non-convex (including trilinear and trigonometric terms), cannot model zero-injection nodes implicitly, and cannot include multi-period constraints without loss of generality.
Another approach \cite{Jovicic2020} proposed a linear optimization approach for joint parameter-state estimation. 
The method \cite{Jovicic2020} is a two-step algorithm, wherein a relaxed ACSE is followed by solving an overdetermined system of injection currents, bus voltages, and unknown parameters with the ordinary least squares (OLS) method to recover the unknown parameters. 
OLS method in the latter step doesn't guarantee the exact solution to the relaxation and can result in parameter estimates that do not satisfy the ACSE constraints.
More recently \cite{Vanin2023} proposed a joint parameter-state estimator for distribution networks.
The method formulates the network constraints with current-voltage (IV) variables and applies a nonlinear programming (NLP) based  Newton-Raphson (NR) solution approach to overcome convergence challenges of the Gauss-Newton-based Weighted Least Squares (WLS) and OLS approaches.
While the method has demonstrated practical empirical results, due to the non-convex (and unobservable) nature of the problem, the solution quality cannot be quantified when the parameters are \textit{truly} unknown.

Dynamic estimation techniques can also recover unknown parameters and provide benefits over techniques that operate on steady-state equations.
For instance, \cite{Zhao2023} and \cite{Costa2022} apply Kalman Filter (KF) and Particle Filter (PF) techniques to estimate unknown grid parameters.
However, these methods require phasor measurement units (PMUs) for high-time-density readings and multi-scan sampling.
The setup is not realizable today, as the majority of grid sensors in the bulk power systems (BPS) are remote-terminal units (RTUs).

Therefore, joint parameter-state estimation techniques based on steady-state equations are most applicable today for estimating grid state and unknown parameters. 
However, in almost all cases, they are formulated as a non-convex optimization.
As such, gauging the solution quality remains a challenge, and convergence to local minima and saddle points is possible as mentioned in literature \cite{weng2016robust}\cite{zhang2019real}(for instance, see Fig. \ref{fig:rand_188_169}), nullifying the practicality of the approach.
Convex relaxations \cite{sdp_acse, conic_acse} can provide a certificate of optimality for the solution, but they deviate from the physics of AC models.

In essence, the fundamental challenge is that of a tradeoff between solution optimality and AC feasibility for non-convex joint parameter-state estimation algorithms.
Furthermore, the underlying problem is inherently \textit{globally} unobservable as many-to-one mappings in a noiseless environment are possible due to nonlinear AC constraints.
Therefore, we will develop a novel circuit-theoretic approach with IV variables to address and mitigate these challenges, effectively.

\subsection{Insights of the problem}

\noindent We formulate a novel circuit-theoretic joint parameter-state estimation problem (ckt-PSE) to address the challenges with the state-of-the-art \cite{Abur2004}, \cite{Jovicic2020}, \cite{Vanin2023}, \cite{Zhao2023}, \cite{Costa2022}. Our novelty is on two fronts:
\begin{enumerate}
    \item First, from the viewpoint of circuit-based network constraints \cite{CircuitTheoretic2020}, we formulate a joint parameter-state estimation with current-voltage (IV) variables; we apply feature transformation to reduce the nonlinearities in ckt-PSE to only bilinear terms.
    \item Second, we apply McCormick relaxation with sequential bound tightening (SBT), based on the foundational work in \cite{Chen2015btACOPF} and \cite{coffrin2015strengthening}  we extend the algorithm in \cite{Nagarajan2016} to ckt-PSE, to provide quantitative guarantees on the estimated solution quality.
\end{enumerate} 

In the results, we demonstrate the efficacy of ckt-PSE for networks up to 2869 nodes. We show that we can recover unknown parameters with quantifiable accuracy. We compare the developed NLP and relaxed convex estimation formulations and consider multi-period constraints to improve solution quality.

\section{Preliminaries}

\subsection{Equivalent Circuit-Theoretic Formulation for AC State Estimation (ckt-SE)} \label{sec:ckt-SE}

\noindent This paper will build on \textit{foundational} equivalent circuit-based approaches for ACSE \cite{CircuitTheoretic2020, jovicic2020enhanced} that solve physics-based Maximum Likelihood Estimation (MLE) and is compatible with both RTUs and PMUs measurements. 
MLE is an observation-driven method. 
It takes observed measurements as apriori information to estimate realizations of random variables from assumed distributions (Gaussian, in this case, an assumption in \cite{Schweppe1970} and applied in \cite{Abur2004}\cite{Jovicic2020linear} and abundance of other works). 
Both \cite{CircuitTheoretic2020, jovicic2020enhanced} aim to find a system state that maximizes the posterior. The methods differ in how they map measurement data to circuit models. 
In a similar vein, these methods also differ from traditional WLS \cite{Schweppe1970}.

We build on this approach \cite{CircuitTheoretic2020} because the underlying circuit model is linear with known convergence guarantees. In this paper, we term this approach ckt-SE. 
The approach uses circuit-based 'IV' instead of the conventional 'PQV' formulation for measurement functions. 
The state variables are real $V_r$ and imaginary $V_i$ voltages instead of the voltage magnitude ($|V|$) and the angle $\theta$. 

\begin{figure}[h]
    \centering
    \includegraphics[scale=0.22]{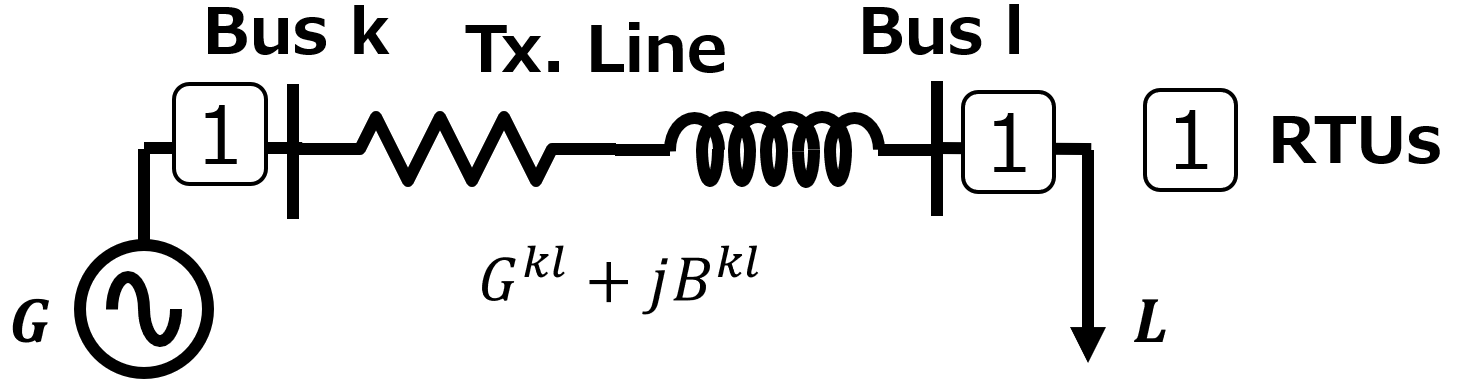}
    \caption{Simple 2-bus network.}
    \label{fig:2bus_network}
\end{figure}

\begin{figure}[h]
    \centering
    \includegraphics[scale=0.2]{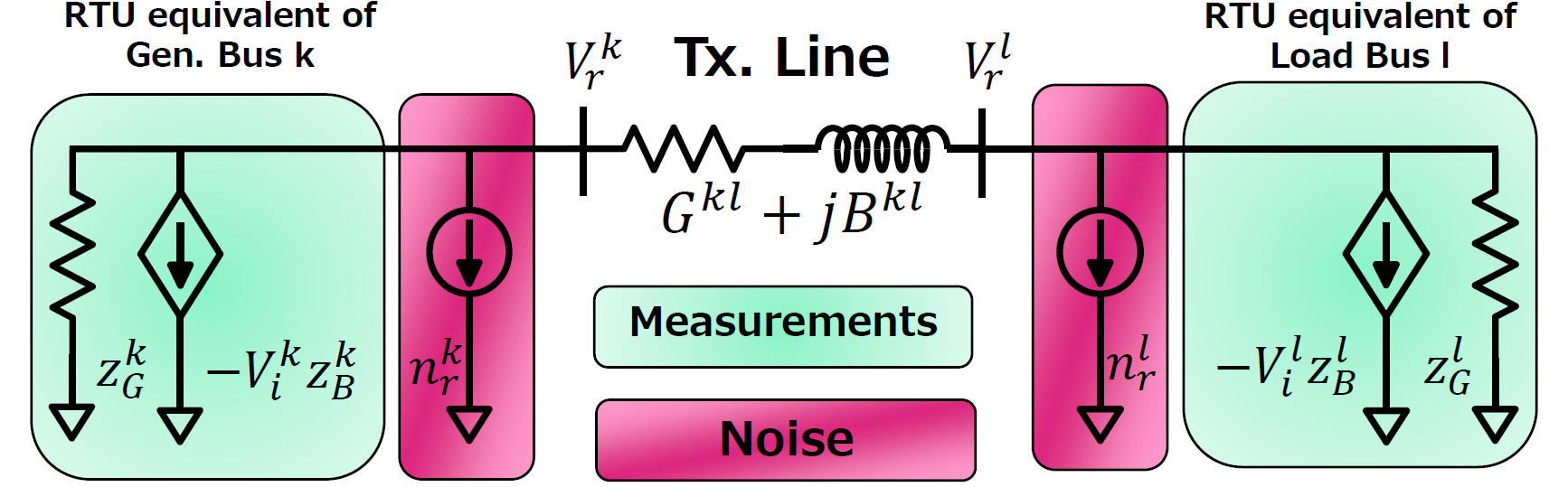}
    \caption{Real sub-circuit of injection measurements-mapped equivalent circuit for the network in Fig. \ref{fig:2bus_network}. The imaginary sub-circuit has a similar structure.}
    \label{fig:2bus_without_params}
\end{figure}

\noindent To illustrate, ckt-SE, consider a simple 2-bus network in Fig. \ref{fig:2bus_without_params}. 
We assume there is an RTU at each of the two buses, measuring voltage magnitude, nodal power injection, and power flow in adjacent line. 
In ckt-SE, we replace the measured injection parameters $(z_P$, $z_Q$, and $z_{|V|})$ with equivalent circuit elements. 
The approach is built on certain relaxations. 
To construct the measurement circuit for RTU nodal power injection, it applies a feature transform to reduce three-dimensional measurements $z_P$, $z_Q$, and $z_{|V|}$ to two dimensions:

\begin{align}
    z_G = \frac{z_P}{{\left(z_{|V|}\right)}^2}; z_B =\frac{z_Q}{{\left(z_{|V|}\right)}^2}
\end{align}

\noindent With the transformation, the nodal network constraints at bus $k$ and $l$ are affine as the injection currents $I^b_{RTU}$ at bus $b \in \{k,l\}$ are linear functions of unknown voltages per KCL:

 \begin{align} \label{eq:rtu_injection}
     I^b_{RTU,r} = z^b_G V^b_r - z^b_B V^b_i + n^b_r\\
     I^b_{RTU,i} = z^b_G V^b_i + z^b_B V^b_r + n^b_i \label{eq:imag_kcl_rtu}
 \end{align}

\noindent Note that \eqref{eq:rtu_injection} includes noise terms $n$ in the measurement circuits as measurements are imperfect or noisy. 
The relaxation is also in the modeling of noise $n$. While traditional WLS-formulation \cite{Schweppe1970} does not assume measurement noise to be Gaussian, traditional MLE-formulation \cite{Abur2004} assumes all measurement noise ($n^z_P, n^z_Q$, and $n^z_{|V|}$) to be Gaussian. 
\cite{CircuitTheoretic2020} assumes the injection current noise to be Gaussian to maintain the key MLE principle. RTUs use wattmeters that measure power through analog or digital processing of voltage and current measurements with averaging, and as such, the Gaussian assumption in \cite{Abur2004} is primarily to preserve key properties of MLE.
 
\begin{figure}[h]
    \centering
    \includegraphics[scale=0.18]{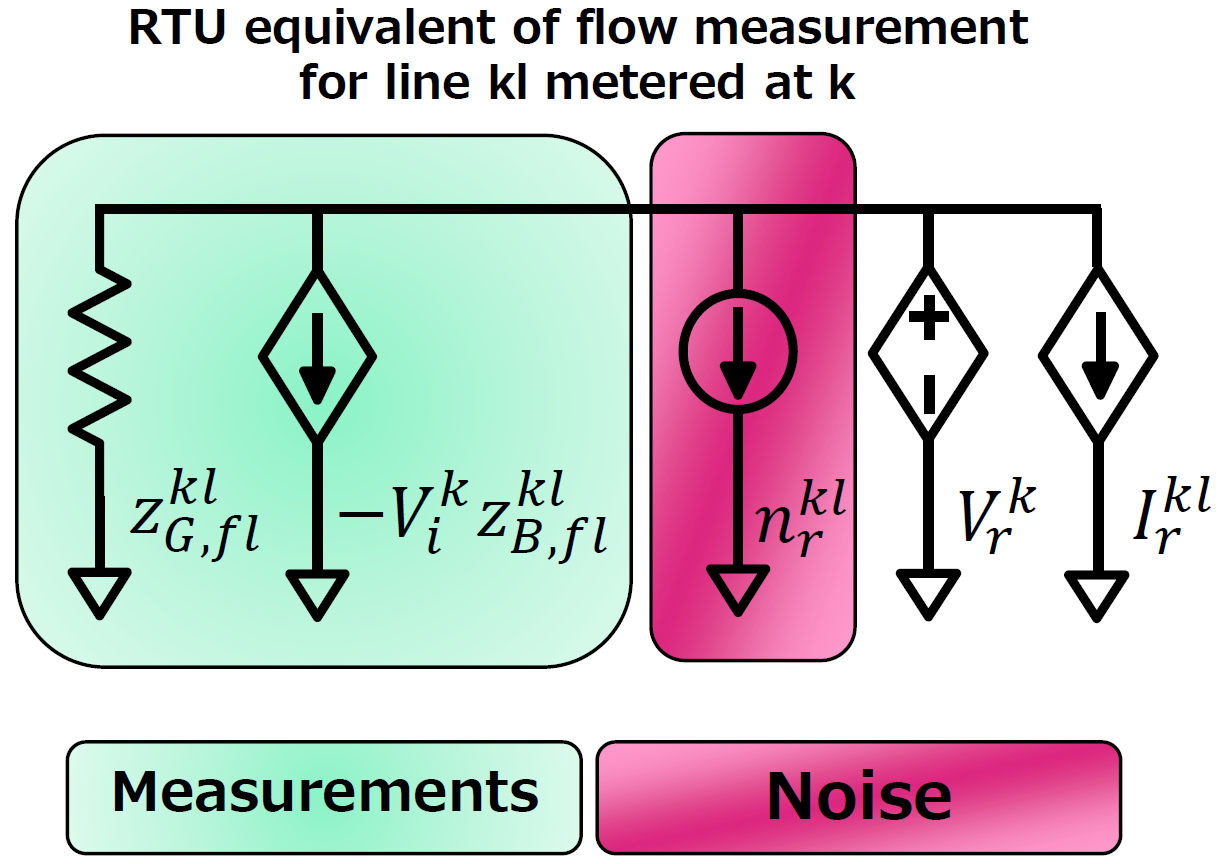}
    \caption{Real sub-circuit of a line flow measurement-mapped equivalent circuit. The voltage $V^k_r$ and $V^k_i$ represent the real and imaginary part of the adjacent bus voltage at bus k, $I_r^{kl}$ is explicitly represented by ECF branch models. The imaginary circuit has a similar structure.}
    \label{fig:line_flow_ECF}
\end{figure}

We can similarly construct the measurement circuit for RTU line flow measurements (see Fig. \ref{fig:line_flow_ECF}). The  measurement function for the line flow measurement between $k$ and $l$, metered at $k$, is given by:

\begin{align}
    z_{G,fl}^{kl}.V_r^k - z_{B,fl}^{kl}.V_i^k + n_r^{kl} + G^{kl}V_r^{kl} - B^{kl}V_i^{kl} = 0\\
    z_{G,fl}^{kl}.V_i^k + z_{B,fl}^{kl}.V_r^k + n_i^{kl} + G^{kl}V_i^{kl} + B^{kl}V_r^{kl} = 0
\end{align}

\noindent where, the flow measurements $z_{P,fl}, z_{Q,fl}$ are relaxed as $z_{G,fl}, z_{B,fl}$:

\begin{align}
    z_{G,fl} = \frac{z_{P,fl}}{{\left(z_{|V|}\right)}^2}; z_{B,fl} =\frac{z_{Q,fl}}{{\left(z_{|V|}\right)}^2}
\end{align}

To solve the ckt-SE, first, an overall circuit for the grid is constructed (see Fig. \ref{fig:2bus_without_params}), replacing any measured component with an equivalent measurement circuit \cite{CircuitTheoretic2020}, \cite{Jovicic2019}. 
Next, to obtain an MLE solution, \cite{CircuitTheoretic2020} minimizes the norm-2 of the noise terms subject to the KCL-based network constraints of the equivalent circuit.

\noindent Mathematically, the problem can be defined as follows:

\begin{subequations} \label{opt:cktSE}
\begin{equation}
        \mathbf{{P}_{cktSE}}: \min_{v,n} f(\textbf{n})\\
\end{equation}
subject to:
\begin{equation}\label{eq:cktSE_hrtu}
         h^k_{RTU}(\textbf{v},\textbf{n}, \textbf{z}) = 0 \quad \forall k \in \mathcal{N}\backslash ZI\\
\end{equation}
\begin{equation} \label{eq:nonlin_pse2}
         h^e_{RTU,fl}(\textbf{v},\textbf{n},\textbf{z}) = 0 \quad \forall e \in \mathcal{E}\\
\end{equation}
\begin{equation}\label{eq:cktSE_hZI}
        h^k_{ZI}(\textbf{v}) = 0 \quad \forall k \in ZI \\
\end{equation}
where the norm-2 minimization in objective has the following form:
\begin{equation}
    f(\textbf{n}) = \min_{v, n} \left( \sum_{k \in \mathcal{I}}  \Big({(n^k_r)}^2 + {(n^k_i)}^2 \Big) + \sum_{ e \in \mathcal{E} } \Big({(n^e_r)}^2 + {(n^e_i)}^2 \Big) \right)\\
\end{equation}
\end{subequations}

\noindent where $\mathcal{I}$ is the set of injection measurements and $ZI$ is the set of zero-injection buses. $\mathcal{E}$ is the set of line flow measurements. 
$\textbf{v}$ is the vector of real and imaginary voltages and $\textbf{n}$ is the vector of noise terms. 
Note that \textit{affine} function $h$ represents IV-based nodal constraints for the equivalent circuit and consists of both measurement parameters ($\textbf{z}$) and other parameters (e.g., line and transformer impedances). \textit{Affine} flow measurement functions is included without loss of generality following \cite{CircuitTheoretic2020}.

Because the constraint set in \eqref{opt:cktSE} is affine, the underlying optimization problem is convex.
It represents a domain-knowledge-driven MLE technique, which has a closed-form result.
Furthermore, including zero-injection nodes naturally shrinks the feasible space to give a more physically meaningful estimation. 


\subsection{McCormick Convex Relaxation}
\noindent The proposed method in \cite{CircuitTheoretic2020} is originally a non-convex formulation with no guarantees on the solution quality. 
As such, we will apply a convex relaxation technique to quantify the \textit{goodness} of solution quality.
As the problem nonlinearities in our approach are bilinear, we will utilize McCormick relaxation, which was proposed in \cite{McCormick1976} and is widely applied in other grid optimization problems like the optimal power flow (OPF) \cite{Javadi2021} and \cite{Bynum2018}.

The McCormick relaxation lifts all bilinear terms in the target equality constraints and represents them with a new lifted variable as in \eqref{eq:Mc_bilinear_term}.
Subsequently, it defines the upper and lower bounds for the variables in the bilinear terms as shown in \eqref{eq:Mc_variable_bounds}. 
With upper and lower bounds defined, it replaces the individual bilinear terms with the set of inequality constraints \eqref{eq:Mc_inequlities}, also referred to as \textit{McCormick Envelope}. 

With this relaxation, the original nonlinear feasible space shown via the green curve in Fig. \ref{fig:SBT_diamond_example} is relaxed to a convex feasible set described via the area inside the boundary of the original bounds in purple. This relaxed feasible space depends on the bounds in \eqref{eq:Mc_variable_bounds}. Thus, the quality of such relaxation critically hinges on the careful and appropriate selection of the upper and lower bounds (see \cite{Javadi2021} for details).

\begin{subequations}\label{eq:McCormick}
\begin{equation}\label{eq:Mc_bilinear_term}
    \begin{split}
    \widehat{x}_{ij} = x_ix_j\\
    \end{split}
\end{equation}
\begin{equation}\label{eq:Mc_variable_bounds}
    \begin{split}
    \underline{x_i} \leq x_i \leq \overline{x_i}\\
    \underline{x_j} \leq x_j \leq \overline{x_j}\\
    \end{split}
\end{equation}
\begin{equation}\label{eq:Mc_inequlities}
    \begin{split}
    \widehat{x}_{ij} \geq \underline{x_i}x_j + x_i\underline{x_j} - \underline{x_i}\underline{x_j}\\
    \widehat{x}_{ij} \geq \overline{x_i}x_j + x_i\overline{x_j} - \overline{x_i}\overline{x_j}\\
    \widehat{x}_{ij} \leq \underline{x_i}x_j + x_i\overline{x_j} - \underline{x_i}\overline{x_j}\\
    \widehat{x}_{ij} \leq \overline{x_i}x_j + x_i\underline{x_j} - \overline{x_i}\underline{x_j}\\
    \end{split}
\end{equation}
\end{subequations}

\subsection{Sequential Bound Tightening (SBT)}

\begin{figure}[h]
    \centering
    \includegraphics[scale=0.15]{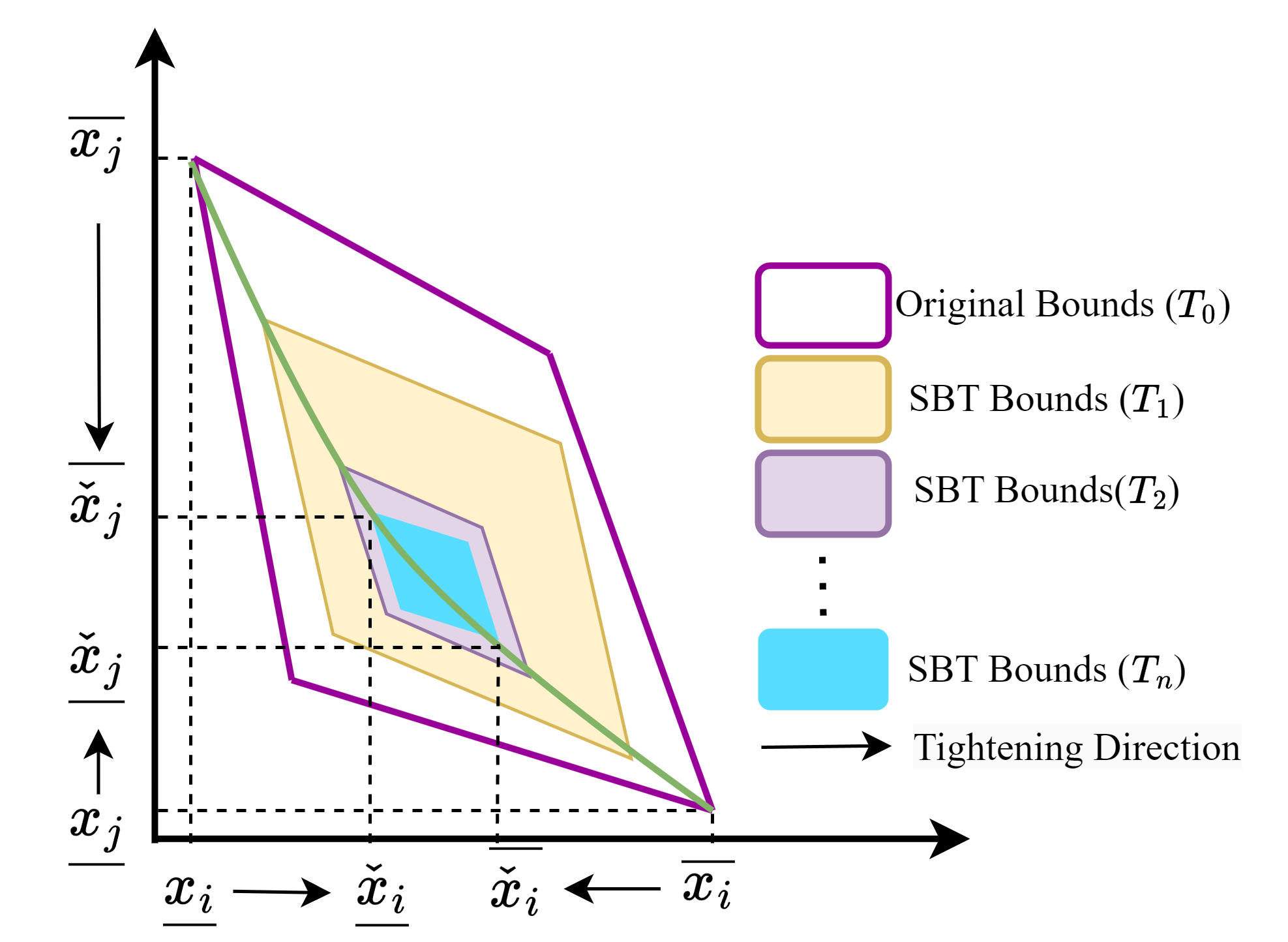}
    \caption{Tightening McCormick Bounds iteratively. From $T_0$ to $T_n$ SBT takes n iterations to find the tightened bounds. $x_i$ shows the original bounds while $\check{x}_i$ shows the tightened bounds. The arrow indicates the tightened direction during the SBT process.}
    \label{fig:SBT_diamond_example}
\end{figure}

\noindent McCormick relaxation guarantees solution optimality but with a trade-off: loss of AC feasibility. 
The quality of upper and lower bounds used to construct the McCormick envelopes quantifies this trade-off in feasibility \cite{Nagarajan2016}. 
For McCormick relaxation to perform better (i.e., provide optimal solutions without deviating significantly from the feasible region), we should seek the tightest possible bounds. 
Fig. \ref{fig:SBT_diamond_example} shows how the feasible space constricts when the bounds at two dimensions tighten to a quarter of their original values. 
In practice, the challenge lies in obtaining tightened bounds that contain the \textit{global optimal} solution of the original NLP problem.

We build upon earlier sequential bound technique (SBT) methods proposed in \cite{Chen2015btACOPF, coffrin2015strengthening, Nagarajan2016}. We specifically extend and apply the SBT algorithm proposed in \cite{Nagarajan2016} to McCormick implementation in this paper to achieve tightened bounds.
When there is no apriori knowledge of good boundaries for a variable, this method can provide tight bounds. 
If there is apriori knowledge of good bounds for a variable, the improvement brought by applying SBT (or other similar method) is limited.

SBT technique inputs the best-known solution of the NLP problem with bilinear nonlinearities $\langle x_i x_j \rangle$ in the equality constraints and the \textit{best-known} upper and lower bounds for the bilinear terms $ \overline{x_i},\overline{x_j}$ and $   \underline{x_i},\underline{x_j}$. It outputs tightened bounds $ \overline{\Check{x_i}},\overline{\Check{x_j}}$ and $   \underline{\Check{x_i}},\underline{\Check{x_j}}$ through an iterative process of optimizing the upper and lower bounds for all variables in the bilinear terms till the improvement reaches a set tolerance as shown in the Fig. \ref{fig:SBT_diamond_example}.

This approach of tightening bounds for each variable in the bilinear term can be computationally prohibitive, considering the number of parallel iterative optimizations we would have to perform.  Therefore, in this work, we will tighten the bounds for only a subset of variables based on domain knowledge.

\section{ckt-PSE: Problem description}
\noindent Next, we design and formulate the circuit-theoretic joint parameter and state estimation (ckt-PSE) problem as an extension of the circuit-theoretic state estimation (ckt-SE). 
The approach estimates the system's state while also estimating the unknown parameters. The parameters of interest include unknown or erroneous network parameters, including line and transformer impedances, and shunt positions. 
We show that the extension is non-trivial, transforming the original convex problem into a non-convex problem. 
We develop the methodology to obtain solutions with minimal trade-offs between solution optimality and AC feasibility. 

\subsection{Circuit-theoretic joint parameter-state estimation formulation (ckt-PSE)}\label{Sec: PSE}

\begin{figure}[h]
    \centering
    \includegraphics[scale=0.2]{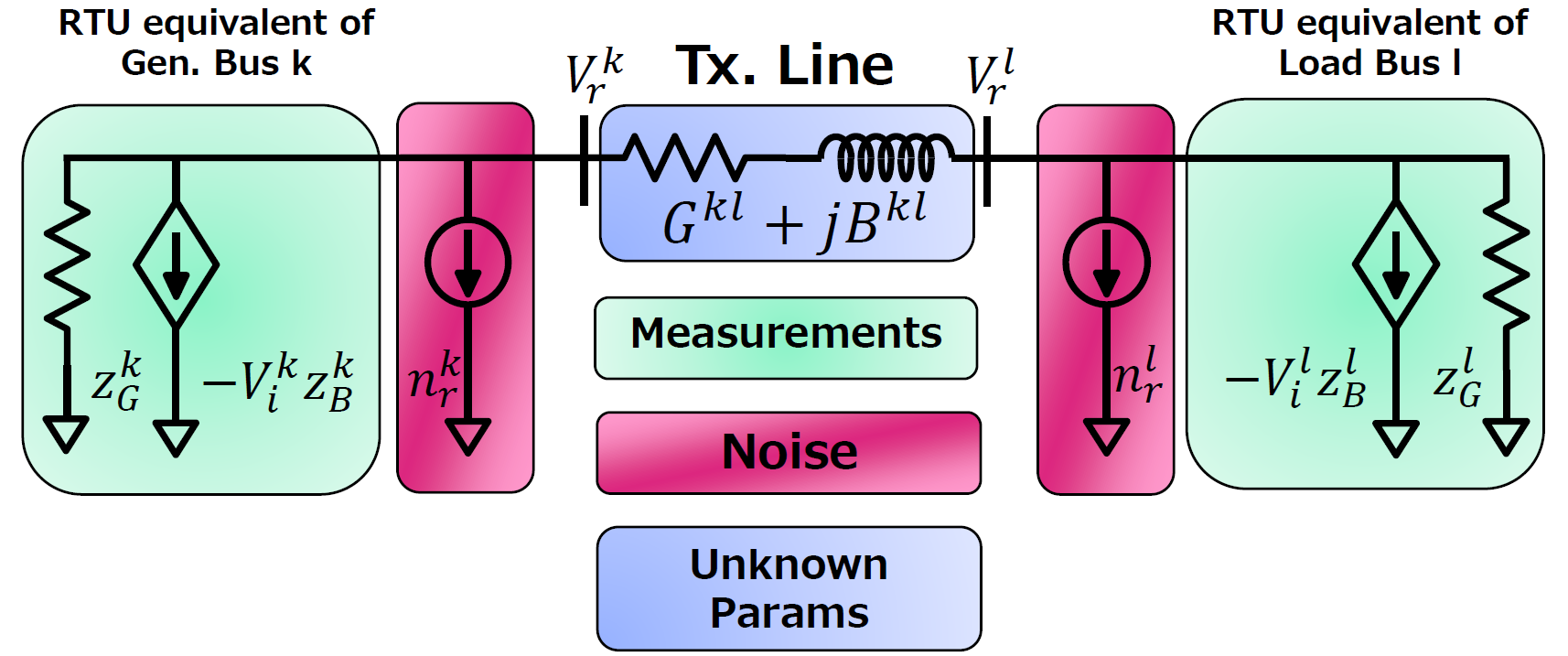} 
    \caption{Real sub-circuit of the grid model for ckt-PSE. Imaginary sub-circuit has a similar structure}
    \label{fig:2bus_with_unknownparams}
\end{figure}

\noindent Here, we propose and formulate a novel \textbf{circuit-theoretic formulation for joint AC parameter-state estimation (ckt-PSE)}. Akin to ckt-SE, in this approach, we first construct an equivalent circuit model of the power grid with measurements mapped as sub-circuits. However, unlike ckt-SE (see Fig. \ref{fig:2bus_without_params}), where all parameters are assumed known and accurate, in ckt-PSE, we operate under the assumption that \textit{certain} or \textit{all} parameters as unknown or erroneous (see Fig. \ref{fig:2bus_with_unknownparams}). 

Therefore, to estimate values of erroneous or unknown parameters, we consider them as variables.
In a manner of state-augmentation, the set of unknown parameters $\boldsymbol{\mathcal{P}}$ are included in the state variable set, resulting in a new expanded variable set \textbf{x}, where \textbf{x} = \textbf{v} $\cup$ $\boldsymbol{\mathcal{P}} \in \mathbb{R}^{|\textbf{v} \cup \mathcal{P}|}$. This will be the new variable set for ckt-PSE. Note that in ckt-PSE, while measurement noise $\textbf{n}$ represents a realization of random variables, the $\boldsymbol{\mathcal{P}}$ is only a set of unknown parameters, not random variables.

 We formulate the ckt-PSE optimization problem as $\mathbf{{P}_{cktPSE}}$ in \eqref{opt:ckt_pse}. We find that the very existence of parameters as variables turns the convex $\mathbf{{P}_{cktSE}}$ into non-convex $\mathbf{{P}_{cktPSE}}$. The nonconvexity stems from converting affine equality constraints in \eqref{eq:cktSE_hrtu}-\eqref{eq:cktSE_hZI} to those with bilinear terms in \eqref{eq:nonlin_pse1}, \eqref{eq:nonlin_pse2}, and \eqref{eq:nonlin_pse3}. We denote the bilinear terms with $\textbf{s} \in \mathcal{B}$.

\noindent In its most na\"ive form, the ckt-PSE has the following form:

\begin{subequations} \label{opt:ckt_pse}
\begin{equation}
   \mathbf{{P}_{cktPSE}}: \min_{\textbf{x},\textbf{n},\textbf{s}} f(\textbf{n})\\
\end{equation}
    subject to:
\begin{equation} \label{eq:nonlin_pse1}
         h^k_{RTU}(\textbf{x},\textbf{n},\textbf{z},\textbf{s}) = 0 \quad \forall k \in \mathcal{N}\backslash ZI\\
\end{equation}
\begin{equation} \label{eq:nonlin_pse2}
         h^e_{RTU,fl}(\textbf{x},\textbf{n},\textbf{z},\textbf{s}) = 0 \quad \forall e \in \mathcal{E}\\
\end{equation}
\begin{equation} \label{eq:nonlin_pse3}
        h^k_{ZI}(\textbf{x},\textbf{s}) = 0 \quad \forall k \in ZI \\
\end{equation}
\begin{equation} \label{eq:binary_terms}
        s^{k} = v^{i} \mathcal{P}^{j} \quad \forall k \in \mathcal{B}\\
\end{equation}
The norm-2 minimization in objective has the following form, with $\mathcal{I}$ is the set of  injection measurements, $\mathcal{E}$ is the set of line flow measurements:
\begin{equation}
    f(\textbf{n}) = \sum_{k \in \mathcal{I}} w_I^k\Big((n^k_r)^2 + (n^k_i)^2 \Big) + \sum_{e \in \mathcal{E}} w_I^e\Big((n^e_r)^2+(n^e_i)^2 \Big)\\
\end{equation}
\end{subequations}

\noindent where $w^k_I$ and $w^e_I$ are the weight for RTU injection node measurement $k$ and the weight for line flow measurement $e$. The choice of weights is discussed in Section \ref{sec:weight_section}. The RTU flow measurement functions $h_{RTU,fl}$ are explicitly included in \eqref{opt:ckt_pse}, (also see Fig.\ref{fig:line_flow_ECF} and Section \ref{sec:ckt-SE} for derivations). The only difference now is that certain parameters may be assumed unknown.





As some parameters are unknown variables in ckt-PSE, the feasible set for the underlying optimization is larger than ckt-SE. As such, there is value in constricting the feasible space with any additional system knowledge. Now, recall the feature transformation of RTU measurements. It results in a loss of information due to feature reduction from 3 measurements to 2 measurements when we retrieve $z_G$ and $z_B$ from $z_{P},z_{Q}$ and $z_{|V|}$. The proposed work will include lost $z_{|V|}$ information by having an additional bilinear voltage-based constraint for measured bus $k$:

\begin{align}
     {(V^k_r)}^2 + {(V^k_i)}^2 - \big({z_{|V|}^k-n^k_{|V|}}\big)^2 = 0 \quad  \forall k \in \mathcal{N}\backslash ZI \label{eq:vol_cons}
\end{align}

\noindent with objective of $\mathbf{{P}_{cktPSE}}$ changing to include noise in voltage measurements $n^k_V$:

\begin{equation}\label{eq:obj_PSE}
    \min_{\textbf{x},\textbf{n}, \textbf{s}}  \sum_{k \in \mathcal{I}} \left(w_I^k\Big( (n^k_r)^2+(n^k_i)^2 \Big) + w_V^k(n^k_V)^2\right) + \sum_{e \in \mathcal{E}} w_I^e\Big( (n^e_r)^2+(n^e_i)^2 \Big)\\
\end{equation}

\noindent In \eqref{eq:vol_cons}, the bilinear terms are sorted in \textbf{s} and extra voltage equality constraints are added to the set $h_{RTU}(\textbf{x},\textbf{n},\textbf{z}, \textbf{s})$. 




\subsection{Unknown vs. erroneous parameter values}


\noindent \nppse{} can incorporate any apriori knowledge of parameters into the algorithm as a \textit{best-known} value of the parameter $\hat{\mathcal{P}}$, changing the unknown parameter as:
\begin{align}
    \mathcal{P} = \hat{\mathcal{P}} + d\mathcal{P} 
\end{align}

\noindent With this apriori knowledge, the network constraint for branch between $k$ and $l$ with unknown parameters $G^{kl}$ and $B^{kl}$ will change as follows:
\begin{align}
     I^{kl}_r = (\hat{G}^{kl} + dG^{kl}) V^{kl}_r -(\hat{B}^{kl} + dB^{kl}) V^{kl}_i\\
      I^{kl}_i = (\hat{G}^{kl} + dG^{kl}) V^{kl}_i + (\hat{B}^{kl} + dB^{kl}) V^{kl}_r
 \end{align}
\noindent With parameters' \textit{best-known} values $\hat{\mathcal{P}}$ being constants, the binary terms in \eqref{eq:binary_terms} will be functions of $d\mathcal{P}$.


\subsection{Applying multi-period sampling}

\noindent While the grid states $\textbf{v}$ vary at every time-sample due to changing operating conditions such as changing loads, generation, and topology, the unknown parameters $\boldsymbol{\mathcal{P}}$ generally remain fixed for a \textit{reasonable} period.
Therefore, by including multi-period sampling into the constraint set, especially from widely varying conditions (high loading vs. low loading), we can \textit{squeeze} the feasible space for unknown parameters, improving the quality of parameter estimates. 
The objective for multi-period sampling changes as follows:
\begin{subequations} \label{opt:ckt_pse_mp}
\begin{equation}
   \mathbf{{P}^{MP}_{cktPSE}}: \min_{{\textbf{x}'},{\textbf{n}'},{\textbf{s}'}} \sum_{t \in \mathcal{T}}  f({\textbf{n}'})\\
\end{equation}
The network constraints for multi-period sampling change as follows:
\begin{equation} \label{eq:mp1}
         h^{k,t}_{RTU}({\textbf{x}'},{\textbf{n}'},{\textbf{z}'}, {\textbf{s}'}) = 0 \quad \forall k \in \mathcal{N}\backslash ZI \quad \forall t \in \mathcal{T}\\
\end{equation}
\begin{equation} \label{eq:mp2}
         h^{e,t}_{RTU,fl}({\textbf{x}'},{\textbf{n}'},{\textbf{z}'},{\textbf{s}'}) = 0 \quad \forall e \in \mathcal{E}, \quad \forall t \in \mathcal{T}\\
\end{equation}
\begin{equation} \label{eq:mp3}
        h^{k,t}_{ZI}({\textbf{x}'},{\textbf{s}'}) = 0 \quad \forall k \in ZI, \quad \forall t \in \mathcal{T} \\
\end{equation}
\begin{equation} \label{eq:mp_binary_terms}
        {\textbf{s}'}^{k} = {v}'^i \mathcal{P}^j \quad \forall k \in {\mathcal{B}'}\\
\end{equation}
\end{subequations}

\noindent Note here that $\textbf{x}', \textbf{n}', \textbf{s}'$ and $\textbf{z}'$ imply that these are concatenated vectors over multiple time periods. Also, that parameters $\boldsymbol{\mathcal{P}}$ remain fixed in the time-period set $\mathcal{T} = \{t_1,...,t_n\}.$

\subsection{Choice of weight} \label{sec:weight_section}
\noindent The source of noise comes from meters' measurement accuracy. We have current measurement noise $w^k_I$ and voltage measurement noise $w^k_V$ in the proposed approach. Commonly the error tolerance is labeled on the meter as $tol: $$\%a + \%b$ or by class. $w^k_I$ and $w^k_V$ should be based on the precision of the equipment to reflect such differences in the objective function: we give larger weight to measurements that have better accuracy and vice versa, to show we trust some accurate meters more in our optimization problem.

\subsection{Observability of the system} \label{Sec:Observability}
\noindent The proposed estimation problem, \nppse{} is unobservable as the underlying system is overdetermined. Therefore, there is no one-to-one mapping between states and measurements in a noiseless scenario \cite{Abur2004} for such a system.

When formulating this problem as an optimization problem, we obtain a nonconvex problem \nppse{}, with many possible solutions, including local minima and saddle points. This is because the constraint set includes nonlinearities. While we cannot enforce one-to-one mapping between states and measurements in a noiseless scenario without relaxing system equations, we strive to deliver \textit{high-quality} estimations by quantifying optimality of the nonconvex \nppse{} problem. 
To that effect, next we formulate and solve variants of $\mathbf{{P}_{cktPSE}}$.

\section{ckt-PSE: Solution Methodology}

\noindent In this paper, we develop three solution approaches for the ckt-PSE problem. 
The first solves the exact NLP problem but without any guarantees of optimality. 
The other two approaches solve the convex relaxation of the problem, balancing network feasibility with optimality.
We show that the final approach is a tight relaxation, achieving optimality while enabling \textit{sufficient} AC model fidelity.

\subsection{Solving ckt-PSE exactly}

\noindent To solve $\mathbf{P_{cktPSE}}$ exactly, we apply primal-dual interior point (PDIP) to \eqref{opt:ckt_pse_mp}. The method ensures high model fidelity by encapsulating AC network constraints; however, because of bilinear terms in the equality constraints, spurious solutions to local minima and saddle points are possible (see Fig. \ref{fig:case8_nlp} in Numerical Results). Thus, next, we explore methods to address this challenge, by exploiting the bilinear nature of the nonlinearities and domain-based bound tightening.




\subsection{McCormick relaxation of ckt-PSE}


\noindent To quantify solution optimality, we relax the bilinear terms with McCormick bounds \cite{McCormick1976} to obtain a lower bound on the NLP. In ckt-PSE, bilinear terms combine real and imaginary node voltage variables $v \in \textbf{v}$ and related unknown parameters $\mathcal{P} \in \boldsymbol{\mathcal{P}}$ as shown in \eqref{opt:ckt_pse}.
By applying McCormick relaxation, each bilinear term introduces a set of inequality constraints as shown in \eqref{eq:Mc_variable_bounds} to \eqref{eq:Mc_inequlities}: $g(\textbf{x},\textbf{s})$. We term the McCormick relaxation of $\mathbf{P_{cktPSE}}$ as $\mathbf{P^{MC}_{cktPSE}}$:


\begin{subequations} \label{eq:mc_cktPSE}
\begin{equation}
    \begin{split}
        \mathbf{P^{MC}_{cktPSE}}: \min_\textbf{x,n,s} f(\textbf{n})\\
    \end{split}
\end{equation}
subject to:
\begin{equation}
    \begin{split}\label{eq:mc_cktPSE_hrtu}
         h^k_{RTU}(\textbf{x},\textbf{n},\textbf{z},\textbf{s}) = 0 \quad \forall k \in \mathcal{N}\backslash ZI\\
    \end{split}
\end{equation}
\begin{equation}
    \begin{split}\label{eq:mc_cktPSE_hZI}
        h^k_{ZI}(\textbf{x},\textbf{s}) = 0 \quad \forall k \in ZI \\
    \end{split}
\end{equation}
\begin{equation} \label{eq:mc_nonlin_pse2}
    \begin{split}
         h^e_{RTU,fl}(\textbf{x},\textbf{n},\textbf{z},\textbf{s}) = 0 \quad \forall e \in \mathcal{E}\\
    \end{split}
\end{equation}
\begin{equation}
    \begin{split}
        g^k(\textbf{x},\textbf{s}) \leq 0 \quad \forall k \in \mathcal{B}\\
    \end{split}
\end{equation}
\end{subequations}

\noindent While McCormick re-formulation guarantees the global optimality of the relaxed problem, the solution quality in terms of AC feasibility (or simply feasibility gap) depends on the bounds used in McCormick envelopes.
Without prior knowledge of parameter values, it is challenging to construct tight McCormick envelopes, and as such, we can obtain solutions that are far from AC feasible solution and lack accuracy as shown in Fig. \ref{fig:case2_MCvsSBT}. 
Therefore, in the following approach, we apply a \textit{modified} sequential bound tightening algorithm, extending the original approach in \cite{Nagarajan2016} to obtain tight bounds for $\mathbf{P^{MC}_{cktPSE}}$.


\subsection{Applying Sequential Bound Tightening (SBT) to ckt-PSE}
\label{sec:apply_SBT}
\begin{figure}[h]
    \centering
    \includegraphics[scale=0.6]{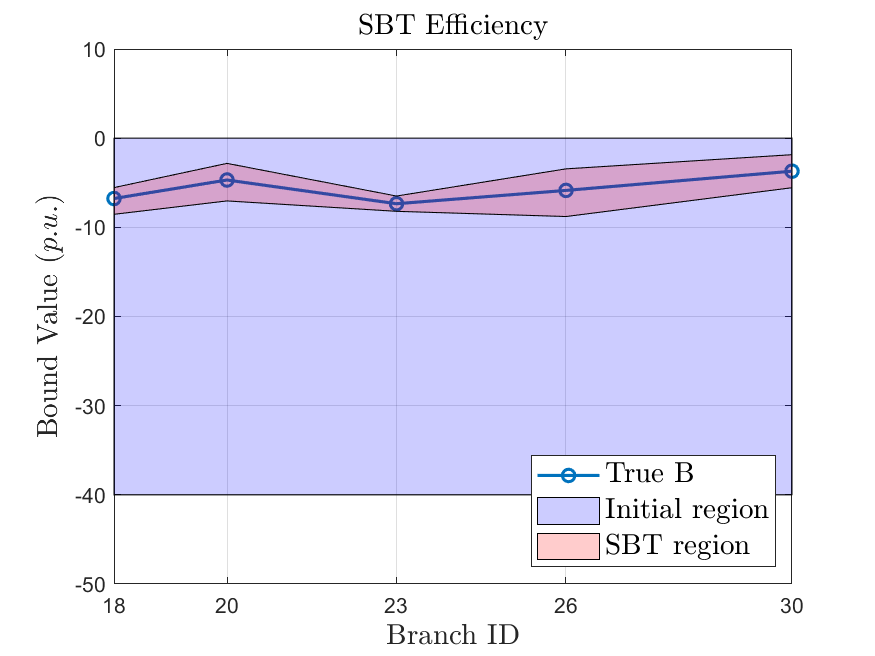} 
    \caption{An illustration of subset-only SBT on IEEE-118 test case with 5 unknown parameters $(jB^{kl})$; The original lower bounds were acquired by using 10 times the true value and upper bound from the definition of $(jB^{kl})$ being negative. Note the polyline connecting each bound is for the sake of showing improvement for each variable corresponding to the given branch on the x-axis, it does not imply any trend between bound and branch.}
    \label{fig:SBT_effective}
\end{figure}

\noindent We extend SBT in \cite{Nagarajan2016} to obtain a set of tight input bounds for \mcpse{}, crucial to creating a good relaxation through McCormick envelopes. The fundamental idea in bound tightening is to transition from \textit{best-known} bounds $(\boldsymbol{\mathcal{P}^L}, \boldsymbol{\mathcal{P}^U})$, to \textit{tightest-feasible} bounds $(\boldsymbol{\check{\mathcal{P}}^L}, \boldsymbol{\check{\mathcal{P}}^U})$. See Fig. \ref{fig:SBT_effective} for a graphical illustration of SBT. To achieve this goal, SBT in \cite{Nagarajan2016} runs parallel batch optimizations to obtain tight bounds by minimizing the gap between bounds of individual variables in bilinear terms. While SBT is originally applied to each variable in bilinear terms set $\textbf{s}$, when the number of unknown parameters and system size grows, this can be computationally prohibitive even though the method is embarrassingly parallel. 

To improve computational performance, we only tighten a subset of the unknown variables: $\boldsymbol{\mathcal{P}}$, where $\boldsymbol{\mathcal{P}} \subset$ \textbf{x}. This is a key difference in our method. We acquire good bounds for bus voltages through the processing of prior SCADA backlog, thus we know voltage state variables inherently have a much tighter range between \textit{best-known} upper and lower bounds. This is in contrast to  \textit{best-known} bounds for unknown parameters where applying SBT is critical. This modification significantly reduces the number of batch optimizations in SBT to just the number of unknown parameters. More importantly, as we only tighten the bounds for unknown parameters, including multi-period constraints results in significantly lower computational overhead as the number of parallel optimizations does not increase for the SBT algorithm, as the same parameter variables are shared across all time periods. Note, however, that the scale of individual optimization will increase due to additional temporal constraints.


\begin{algorithm}
\caption{Modified SBT (Extension of Algorithm in \cite{Nagarajan2016})}
\label{alg:SBT_MINLP}
\begin{algorithmic}[1]
\label{algo:sbt}
\State $\textbf{Input:}$ $\boldsymbol{\mathcal{P}}$, $\boldsymbol{\mathcal{P}^L}$,  $\boldsymbol{\mathcal{P}^U}$, $\boldsymbol{n^*}$ (unknown parameter set, its corresponding best-known bounds, and NLP solution)

\State \textbf{Output:} $\boldsymbol{\check{\mathcal{P}}^L}$, $\boldsymbol{\check{\mathcal{P}}^U}$ (tightened bounds)
\State \textbf{Set} $\epsilon>0$, Set tolerance to greater than 0
\State \textbf{Solve} in parallel $\forall \mathcal{P}^i \in \boldsymbol{\mathcal{P}}$, $i$ $\in \{1,...,|\mathcal{P}|\}$
\Indent
    \State $\textbf{Set}$ $\underline{\mathcal{P}^{i}}$ $\leftarrow$ $\boldsymbol{\mathcal{P}^L}$, $\overline{\mathcal{P}}^{i}$ $\leftarrow$ $\boldsymbol{\mathcal{P}^U}$,$\underline{\mathcal{P}}^{i_{iter}} = \overline{\mathcal{P}}^{i}_{iter} = 0$
    
    \While{$||\underline{\mathcal{P}}^{i} - \underline{\mathcal{P}}^{i}_{iter}||_2 > \epsilon$ and $||\overline{\mathcal{P}}^{i} - \overline{\mathcal{P}}^{i}_{iter}||_2 > \epsilon$}
    
    \State $ \underline{\mathcal{P}}^{i}_{iter} \gets \underline{\mathcal{P}}^{i}$ and $\overline{\mathcal{P}}^{i}_{iter} \gets \overline{\mathcal{P}^{i}}$ 

    \State $\check{\underline{\mathcal{P}}}^{i}:=\min_{\textbf{x},\textbf{n}, \textbf{s}}\underline{\mathcal{P}}^{i}$
    
    \State $\overline{\check{\mathcal{P}}}^{i}:=\max_{\textbf{x}, \textbf{n}, \textbf{s}}\overline{\mathcal{P}}^{i}$
    \Indent
    \State$\textbf{s.t.} f_{\textbf{x,n,s}}(\boldsymbol{n}) \leq f_{\textbf{x,n,s}}(\boldsymbol{n^*})$
    
    \State $h(\textbf{x},\textbf{z}, \textbf{n},\textbf{s}) = 0$
    
    \State $g(\textbf{x},\textbf{s}) \leq 0$
    

    \EndIndent
    
    \State $\underline{\mathcal{P}^{i}}  \gets 
    \check{\underline{\mathcal{P}}^{i}}$ and 
    $\overline{\mathcal{P}^{i}} \gets \overline{\check{\mathcal{P}}^{i}}$ (update $\boldsymbol{\mathcal{P}^L}$ and $\boldsymbol{\mathcal{P}^U}$)
    \EndWhile\\
{All final parameter bounds packed in $\boldsymbol{\check{\mathcal{P}}^U}$ and $\boldsymbol{\check{\mathcal{P}}^L}$} 
\EndIndent
\end{algorithmic}
\end{algorithm}

Algorithm 1 describes the SBT algorithm. We first solve \nppse{} to obtain the lowest available objective for unrelaxed \textbf{NLP} problem $f(\boldsymbol{n^*})$ and add it as a constraint to bound the objective of SBT optimizations in Algorithm 1. While bounding the objective, we run in parallel, an optimization for each parameter's lower and upper bounds. 
The algorithm concludes by providing new tighter bounds to apply to McCormick relaxation $\mathbf{P^{SBT}_{cktPSE}}$. Here, the problem has the same structure as \mcpse{} but with better-defined (tighter) bounds from SBT. Thus, better relaxation performance (smaller variance and lower AC violation) is expected from such a processed problem.

\begin{remark}
    \textit{Both McCormick formulations, with or without SBT, are compatible with multi-period ($\mathbf{{P}^{MP}_{cktPSE}}$), without loss of generality. But as the number of time periods increases, the number of bilinear terms will increase linearly. However, in the extended SBT approach in Algorithm 1, where we perform batch optimizations for tightening the bounds for unknown parameters only, no additional computational overhead for multi-period constraints will incur because the same unknown parameters are shared across all temporal constraints.}
\end{remark}

\begin{remark}
    \textit{SBT can have a large computation burden because the algorithm requires solving many convex optimizations, as many as the number of unknown parameters. However, as we apply SBT to only unknown parameters, which are typically stable and do not change once calculated, the SBT bounds can be used without recalculation for all subsequent estimations until there is a topology change or adjustment to data, such as transformer taps.}
    
\end{remark}

\section{Experimental Setup}
\noindent To demonstrate the efficacy of the proposed algorithms, we will run \nppse{}, \mcpse{}, and \sbtpse{} on four cases. Case 1, which represents a 14-bus network, assumes all branch parameters are unknown. Case 2, an IEEE 118 bus network, better illustrates the effectiveness of \sbtpse{} and the corresponding use of McCormick envelopes. Case 3 represents a larger network with 2869 nodes and demonstrates some weaknesses of the \nppse{} approach. Case 4, also built off the IEEE 118 bus network includes unknown shunt parameters. Details for the cases are given in Table \ref{tab:Case_setup}.

\setlength{\tabcolsep}{6pt}
\begin{table}[]
    \caption{Experiment set up}
    \centering
    \renewcommand{\arraystretch}{1.5} 
    \begin{tabular}{@{}cccc@{}}
        \hline
        \textbf{Scenario} & \textbf{Test Case} & \textbf{Unknown $B$ Id$^{\#}$} & \textbf{Unknown $B_{sh}$ Id$^{\#}$} \\
        \hline
        Case 1 & IEEE-14 & all & NA\\
        Case 2 &  IEEE-118 & $5, 52, 54, 84, 103, 169$ & NA\\
        Case 3 & 2869pegase & $175 ,1725$ & NA\\
        Case 4 & IEEE-118 & $18, 20$ & 4, 25, 37, 90\\
        \hline
        \multicolumn{4}{l}{\footnotesize $\#$ The id is the branch or shunt id in the .raw file for which $B$ is unknown.} \\
    \end{tabular}
    \label{tab:Case_setup}
\end{table}

\subsection{Data creation and Experiment setup}

\noindent For each case, we run 100 instances of parameter-state estimation algorithms (\nppse{}, \mcpse{}, and \sbtpse{}). We assume RTU injection measurements at all nodes, excluding zero-injection nodes, $(\forall k \in \mathcal{N}\backslash ZI)$. We also assume 2 line flow RTU measurements in test case 1 and 10 in test case 2, we chose the location of these measurements arbitrarily. To generate synthetic measurement data for each instance of each case, we run power flow simulations and add noise to the results. The noise samples are obtained based on a standard deviation value of 0.001  \cite{CircuitTheoretic2020} for both current and voltage measurements. For unknown parameters, we assume no prior knowledge of the magnitudes and, therefore, use initial bounds for McCormick envelopes as large as 1000\%. The number is chosen to be larger than what's commonly used in other literature \cite{jovicic2021computationally}.
We are designing the overall experiment to answer the following questions:
\begin{enumerate}
    \item \textbf{Robustness of the NLP approach \nppse{}:} Does the NLP approach generally produce robust and accurate estimates when given a good initial guess? Does the property hold when a good initial guess is unavailable?
    \item \textbf{The improvement of \sbtpse{} compared to \mcpse{}:} Does the use of SBT improve the parameter bounds, and subsequently improve the quality of state and unknown parameter estimates?
    \item \textbf{The scalability of the proposed algorithm:} Do the methods scale when the system size and the number of unknown parameters increase?
\end{enumerate}

\noindent To illustrate our findings and compare the various estimation techniques, we use the following error metrics RMSE, NRMSE and variance ($\sigma^2$): 

\begin{equation}
\begin{split}
       RMSE_{\mathbf{v}} = \sqrt{\frac{\sum^n_{i=0} \sum^{|\mathbf{v}|}_{j=0} ({v}^{i,j}_{est} - {v}^{i,j}_{true})^2}{n|\mathbf{v}|}}
\end{split}
\end{equation}
\begin{equation}
\begin{split}
    NRMSE_{\mathbf{v}} = \frac{RMSE_{\mathbf{v}}}{\overline{v}_{est}}\label{eq:NRMSE}
\end{split}
\end{equation}
\begin{equation}
\begin{split}
    \sigma^2_{v,j} = {\frac{\sum^n_{i=0} ({v}^{i,j}_{est} -\overline{v}_{est}^{i,j})^2}{n}} \quad \forall j \in {1...\mathbf{|v|}}
\end{split}
\end{equation}
\begin{equation}
\begin{split}
    \sigma^2_{v,ave} = \frac{\sum^{|\mathbf{v}|}_{j=0}\sigma^2_{v,j}}{|\mathbf{v}|}
\end{split}
\end{equation}

\noindent where $v_{est}$ is the estimate of state $v$. Note the RMSE and variance for parameters ($\text{NRMSE}_{\mathcal{P}}$ and $\sigma^2_{p,avg}$) are calculated in the same manner as state variables $v$. 

In another metric, percentage estimation error, we compare the estimated value of parameters to the true value. We use this metric to show an absolute comparison of estimates from \nppse{}, \mcpse{}, and \sbtpse{}, and define it for a given estimate $y_{est}$ and its true value $y_{true}$, by:
\begin{align}
    \text{Estimation Error Coefficient}(\%) &= \frac{({y}_{est} - {y_{true})}}{y_{true}} 100
\end{align}

\noindent 
In this experiment, we also assume the same accuracy for various RTU meters (i.e., $\forall k \in \mathcal{R}, w^k_I,w^k_V$ are uniform). The user can choose more realistic weights with access to better precision and tolerance data for meters. 

\noindent \textbf{Experiment Environment and Codebase} \label{sec:experiment_environment}: Our code and data are publicly available at
\text{https://github.com/Asang97/cktPSE.git}. The proposed algorithm is implemented in Python with network data in .raw format. All simulations are processed on a PC with an AMD Core Razen 3 3100 CPU and 16 GB of RAM. We used open-source optimizer \textsc{IPOPT} (version 3.14.10) for all optimizations.

\section{Numerical Results}

\subsection{Accuracy and Robustness}

\noindent Table \ref{tab:Est_accuracy} illustrate the performance of \nppse{}, \mcpse{}, and \sbtpse{} for four (4) different test cases, each of them run with 100 samples of noise. In most cases, the non-convex \nppse{} approach performs best among the three methods. However, the certificate of optimality cannot be obtained for \nppse{}, and when parameters are truly unknown, this can be a challenge, as performance is hard to gauge.

\noindent \textbf{Variance metric for Outliers:} We use both variance and NRMSE as metrics to evaluate the quality of our estimation. Specifically, in case 3 for NLP and cases 2 and 3 for McCormick, high variance shows low-quality estimates in many instances. We observe the high variance is due to the use of wide bounds in McCormick in \mcpse{} and rare convergence to local solution for \nppse{}.
\begin{table}[]
\caption{Estimation Accuracy Evaluation}
    \centering
    \renewcommand{\arraystretch}{1.5} 
    \begin{tabular}{@{}ccccc@{}}
    \hline
    \textbf{Algorithm/Case} & $\textbf{NRMSE}_{\textbf{v}}$ &$\sigma^2_{v,avg}$&$\textbf{NRMSE}_{\mathcal{P}}$&$\sigma^2_{\mathcal{P},avg}$\\ \hline
    \nppse{}/1          & 6.30E-04 & 9.50E-06 & 0.05     & 0.12     \\
    \mcpse{}/1           & 8.60E-03 & 7.10E-05 & 0.14     & 0.02     \\
    \sbtpse{}/1       & 9.65E-04 & 9.26E-07 & 0.03     & 0.05 \\ \hline
     \nppse{}/2          & 2.10E-04 & 8.40E-05 & 1.50E-05 & 3.30E-06 \\
    \mcpse{}/2           & 1.10E-03 & 8.50E-05 & 7.70E-02 & \textcolor{blue}{11.08}\\
    \sbtpse{}/2       & 7.50E-03 & 1.30E-04 & 3.40E-02 & 1.55     \\ \hline
     \nppse{}/3          & 3.00E-02 & 8.90E-04 & 0.10      & \textcolor{red}{69.37}\\
    \mcpse{}/3           & 2.80E-02 & 5.70E-04 & 0.12     & \textcolor{blue}{15.45}\\
    \sbtpse{}/3       & 1.90E-02 & 1.20E-05 & 0.07     & 2.20E-06 \\ \hline
     \nppse{}/4          & 2.40E-04 & 8.40E-03 & 6.70E-03 & 1.30E-04 \\
    \mcpse{}/4           & 1.60E-03 & 8.60E-05 & 0.21     & 1.10E-03 \\
   \sbtpse{}/4       & 1.10E-03 & 8.50E-05 & 0.07     & 1.80E-03 \\ 
   \hline
\end{tabular}\label{tab:Est_accuracy}
\begin{minipage}{9cm}
\vspace{0.1cm}
\footnotesize 1. Results in \textcolor{blue}{blue} indicate higher bad parameter estimate instances due to wide McCormick bounds\\
\footnotesize 2. Results in \textcolor{red}{red} indicate a high number of bad parameter estimates due to sub-optimal solutions in NLP
\end{minipage}
\end{table}



\begin{table}[]
\caption{Algorithm Feature Summary}
    \centering
    \renewcommand{\arraystretch}{1.5} 
    \begin{tabular}{@{}ccc@{}}
    \hline
    \textbf{Algorithm} & \textbf{Optimality} & \textbf{AC Feasibility}\\ \hline
    \nppse{}        & $\times$ & $\checkmark$      \\
    \mcpse{}         & $\checkmark$ & $\times$      \\
    \sbtpse{}      & $\checkmark$ & $\times$  \\ \hline
\end{tabular}\label{tab:Optimality_feasibility}
\begin{minipage}{9cm}
\vspace{0.1cm}
    \footnotesize Optimality property describes whether the algorithm is guaranteed to converge to the global optima, and the AC feasibility property describes whether the algorithm satisfies the AC KCL-based network constraints within a certain tolerance.
\end{minipage}
\end{table}

\subsection{SBT tighten efficiency}

\begin{figure}[h]
    \centering
    \includegraphics[scale=0.70]{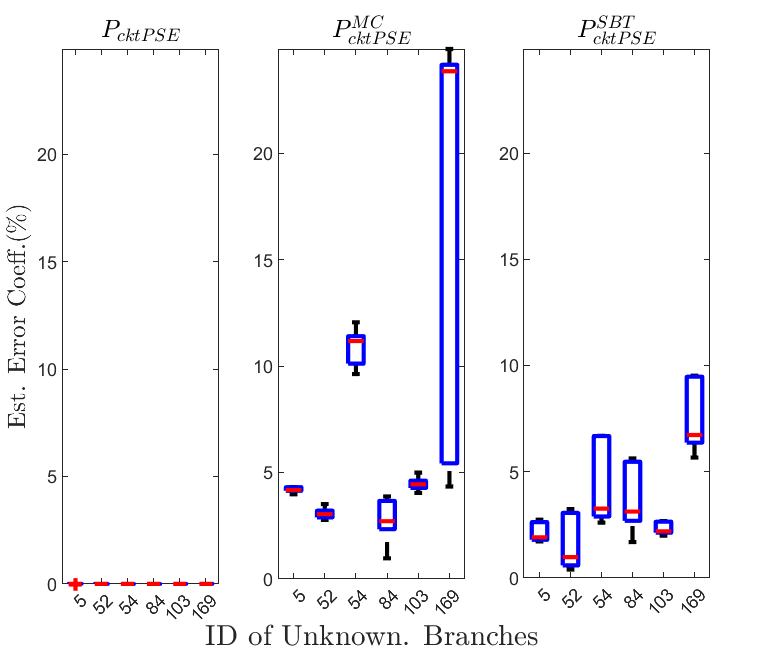}
    \caption{The estimation error comparison between \mcpse{} and \sbtpse{} of case 2 The box indicates the lower and upper quartile of the estimation with a red bar showing its median. Starched whiskers showing maximum and minimum. In general, \sbtpse{} performs better than \mcpse{}. \nppse{} performs the best.}
    \label{fig:case2_MCvsSBT}
\end{figure}

\begin{figure}[h]
    \centering
    \includegraphics[scale=0.33]{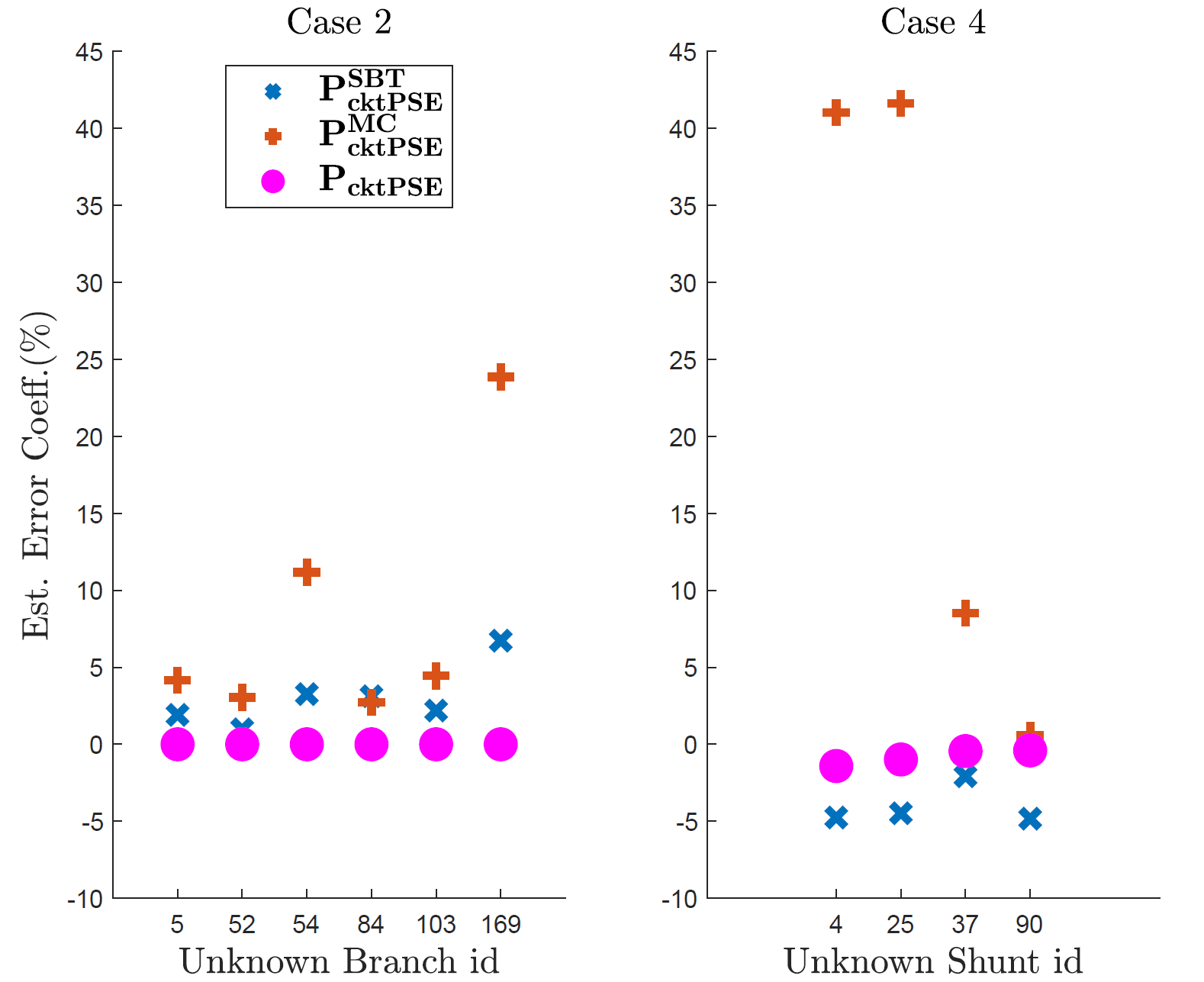} 
    \caption{Method-wise Parameter Estimation Error Comparison for case 2 and 4. Each point is the Error Coefficient between the true value and group median of est. parameters.}
    \label{fig:case2&3_scatter}
\end{figure}

\noindent While \nppse{} does not guarantee solution optimality, on the other hand, McCormick relaxation guarantees optimality but with a loss of AC feasibility. Both Fig. \ref{fig:case2_MCvsSBT} and Fig. \ref{fig:case2&3_scatter} show that the estimation quality deteriorates with the use of \mcpse{}. We posit this is due to using wide bounds for McCormick envelopes. \sbtpse{} addresses this problem by tightening the McCormick bounds. As an example, Fig. \ref{fig:SBT_effective} shows the bound tightening of various branch parameters in case 2 with SBT. We observe the bound gap decreased significantly from the initial region to the SBT region. 

Fig. \ref{fig:case2_MCvsSBT} shows the error coefficients of estimations of case 2. By comparing the median of each box, we see improvement in the error coefficient from \mcpse{} to \sbtpse{} in most parameters.
Fig.\ref{fig:case2&3_scatter} illustrates in both cases, \mcpse{} shows some off-target estimation due to its wide McCormick bounds, and when the bound is tightened, we can observe \sbtpse{} estimation being more accurate while \nppse{} is providing effective estimations. Combining the results shown in table \ref{tab:Est_accuracy}, despite that the \nppse{} guarantees no global optima, it lies close to the true value. When it comes to the convex relaxation, both \mcpse{} and \sbtpse{} results have some offset compared to \nppse{} while in all \sbtpse{} performs better.
\sbtpse{} deliver a better estimation of parameters with a smaller NRMSE and more AC feasible results as shown in Table \ref{tab:Est_accuracy}. 

\begin{figure}[h]
    \centering
    \includegraphics[scale=0.70]{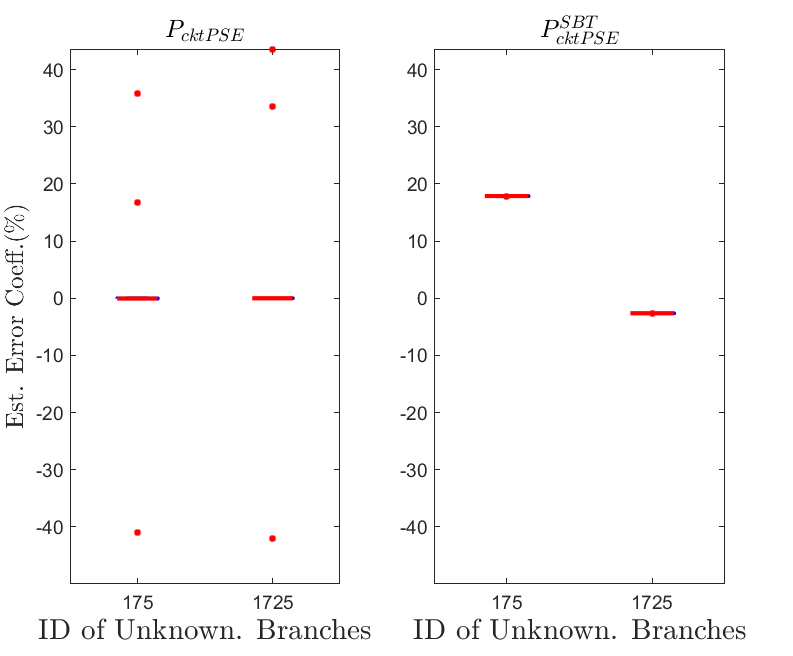}
    \caption{The estimation error comparison between \nppse{} and \sbtpse{} of case 3. The points indicate outliers (due to convergence to local minima or saddle points), and the red bar indicates the group median. The objective values for three outliers in the left subplot range from 64.86 to 278.83; in comparison, the objective values for on-target estimations range from 0.31 to 0.33. This indicates outliers may have resulted from convergence to local optima or saddle points as all noise is randomly sampled from the same non-gross Gaussian distribution.}
    \label{fig:case8_nlp}
\end{figure}

\subsection{Sub-optimal solutions from NLP \nppse{}} \label{local_solutions}

\begin{figure}[h]
\begin{center}
    
    \includegraphics[scale=0.22]{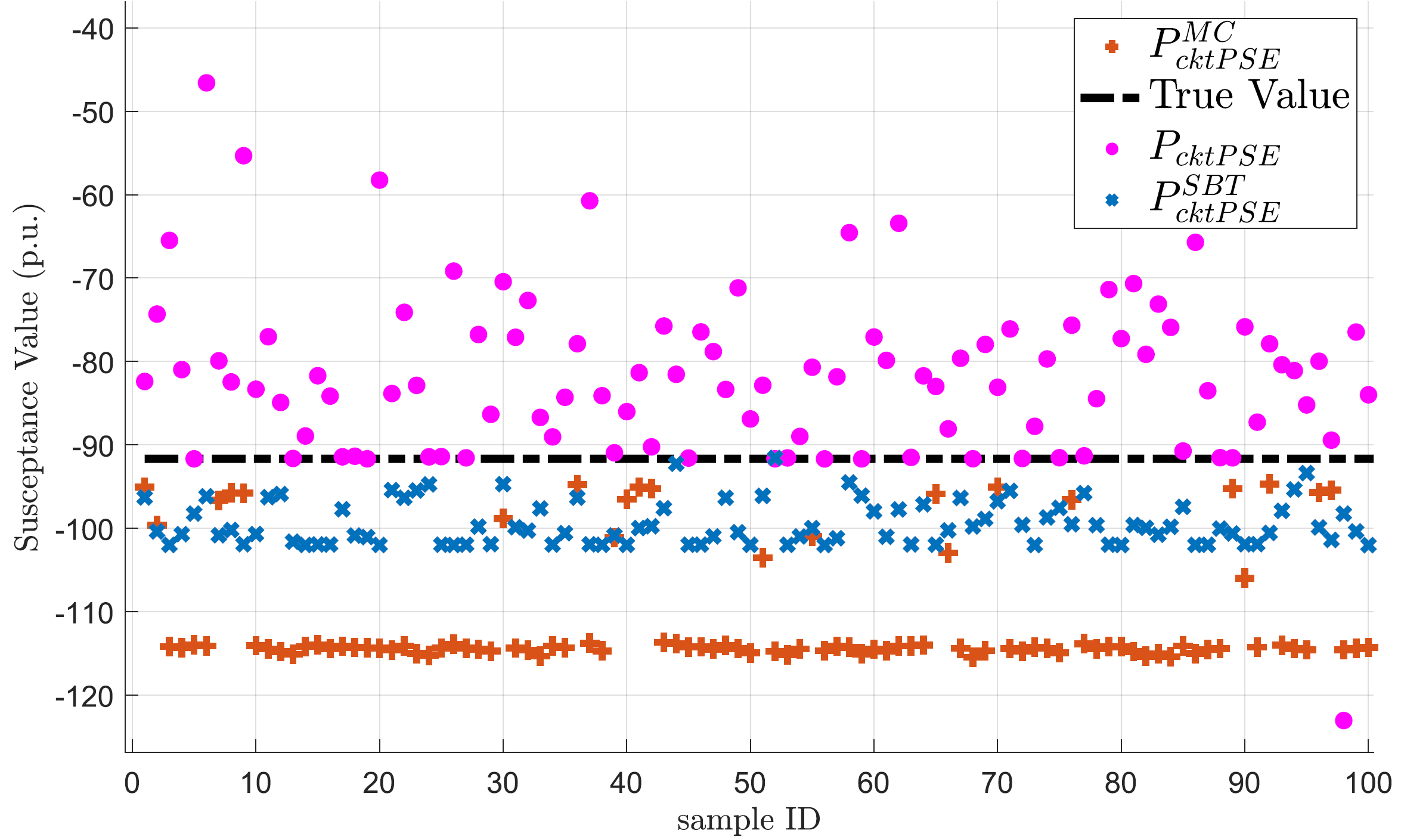}
    \caption{Branch 169 parameter estimate from 100 runs of all 3 cktPSE parameter estimation algorithms. The initial conditions for these algorithms were randomly chosen from $|v_i| \in [0.95 p.u. , 1.05 p.u.]$ $\angle v_i \in [-30^\circ, 30^\circ]$. All other unknown branches have a pattern similar to this figure.}
    \label{fig:rand_188_169}
    \end{center}
\end{figure}

\noindent \nppse{} generally has the best performance when a good initial condition is available (e.g. when the previous time-step estimate is used as an initial guess for the current time-step). Nonetheless, convergence to non-optimal solutions can result in \textit{bad} estimates at times. See, for example, Fig. \ref{fig:case8_nlp}, which shows the estimation quality of two unknown parameters in a 2869-node network. Of 100 samples, \nppse{} generally retrieves the unknown parameters of two lines with almost no error. However, in rare cases, it converges to local minima or saddle points, resulting in \textit{bad} estimates, as observed in the figure's outliers. The likelihood of sub-optimal solutions (i.e., local minima or saddle points) is higher when good initial conditions are unavailable (see Fig. \ref{fig:rand_188_169}). 
 Using the \sbtpse{} algorithm can overcome the challenge with \nppse{}. However, it comes with a tradeoff in terms of increased average error. 

\subsection{Scalability of the proposed algorithm}
\noindent We study the scalability of ckt-PSE by documenting the:
\begin{enumerate}
    \item Solving time for networks with varying numbers of system nodes (2\% parameter unknown)
    \item  Solving time as the number of unknown parameters increases for the 2869 nodes network
\end{enumerate}

\begin{figure}[h]
    \centering
    \includegraphics[scale=0.65]{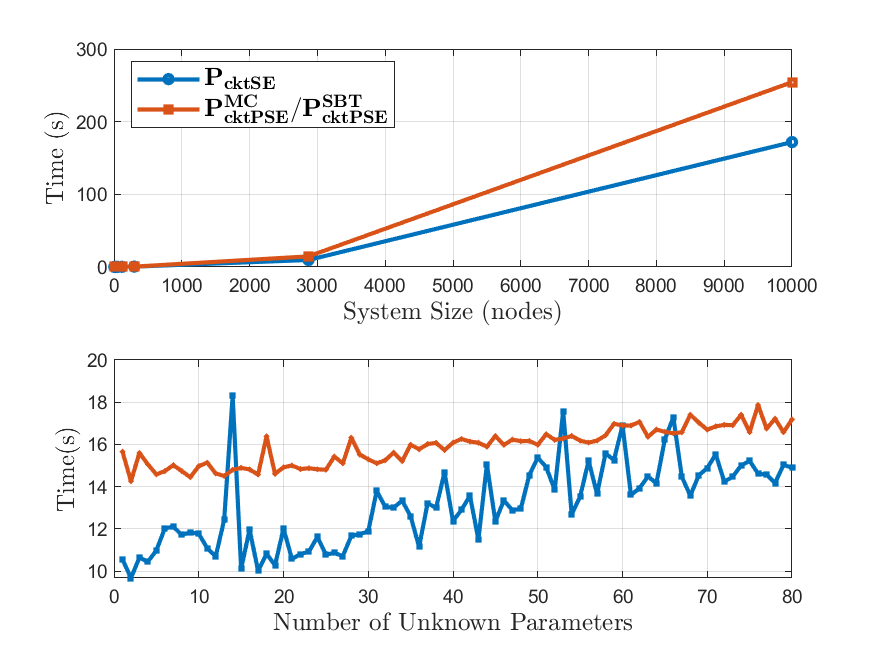}
    \caption{Method scalability in terms of the number of nodes and the number of parameters.}
    \label{fig:run_time}
\end{figure}

\noindent Fig. \ref{fig:run_time} shows the time complexity of algorithms with increasing system size. For the  10k node network (ACTIVSg10k), \nppse{} takes 30 iterations to converge, and \sbtpse{} takes 38 iterations to converge. While for the 2869 node network, \nppse{} takes 16 iterations to converge, and \sbtpse{} takes 25 iterations to converge. 
Note that the computational complexity of bound tightening in SBT is not included in Fig. \ref{fig:run_time} as it is only performed once.
We observe an almost linear increase in simulation time with the increasing number of unknown parameters. 
For the 2869 system, the NLP approach's solution time almost increased by 50\% as the number of unknown parameters grew from 0 to 80. 


\section{Conclusions}
\noindent We developed a novel circuit-theoretic framework for joint parameter-state estimation (ckt-PSE).
Due to the circuit mapping of the grid constraints, ckt-PSE is less nonlinear, including only bilinear terms. 
To balance the trade-off in optimality and AC feasibility (see Table \ref{tab:Optimality_feasibility}), we propose three algorithms for ckt-PSE: an exact NLP (\nppse{}) and convex relaxations (\mcpse{}, \sbtpse{}).
We draw these final conclusions:
\begin{itemize}
    \item \nppse{} approach performs the best in most scenarios when good initial conditions are available (e.g. good estimate from prior time-step); however, it cannot provide the certificate of optimality and exhibits sub-optimal solutions in some scenarios when good initial conditions are unavailable.
    \item \sbtpse{} presents a convex relaxation, an improvement on the naive McCormick approach, presenting a balance between optimality and AC feasibility. 
    \item With domain knowledge, we reduce the problem complexity of the SBT algorithm, by tightening only a subset of bilinear terms (i.e. unknown parameters $\boldsymbol{\mathcal{P}}$)
\end{itemize}
As a final note, our experiments showcase the ability to recover unknown states and parameters with a quantifiable level of accuracy for realistic-sized networks.

\bibliographystyle{IEEEtran}


\end{document}